\documentclass[aps,prl,reprint,superscriptaddress]{revtex4-1}
\usepackage{graphicx}% Include figure files
\usepackage{amsmath}
\usepackage{physics}
\usepackage{color}
\usepackage{notes2bib}
\usepackage{hyperref}
\usepackage{layouts}
\usepackage{bbold}
\usepackage{mathtools}
\begin{document}
% Use the \preprint command to place your local institutional report
% number in the upper righthand corner of the title page in preprint mode.
% Multiple \preprint commands are allowed.
% Use the 'preprintnumbers' class option to override journal defaults
% to display numbers if necessary
%\preprint{}

%Title of paper
\title{Excitation Pathways in Resonant Inelastic X-Ray Scattering of Solids}

\author{Christian Vorwerk} \email[]{vorwerk@physik.hu-berlin.de}
\affiliation{Physics Department and IRIS Adlershof, Humboldt-Universit\"at zu
Berlin, Berlin, Germany} \affiliation{European Theoretical Spectroscopy
Facility (ETSF)}
\author{Francesco \surname{Sottile}} \affiliation{LSI, Ecole Polytechnique,
CNRS, CEA, Institut Polytechnique de Paris, F-91128 Palaiseau, France}
\affiliation{European Theoretical Spectroscopic Facility (ETSF)}
\author{Claudia \surname{Draxl}} \affiliation{Physics Department and IRIS
Adlershof, Humboldt-Universit\"at zu Berlin, Berlin, Germany}
\affiliation{European Theoretical Spectroscopic Facility (ETSF)}

%Collaboration name if desired (requires use of superscriptaddress
%option in \documentclass). \noaffiliation is required (may also be
%used with the \author command).
%\collaboration can be followed by \email, \homepage, \thanks as well.
%\collaboration{}
%\noaffiliation

\date{\today}

\begin{abstract}
We present a novel derivation of resonant inelastic x-ray scattering (RIXS)
which yields an expression for the RIXS cross section in terms of emission
pathways between intermediate and final many-body states.  Thereby electron-hole
interactions are accounted for, as obtained from full diagonalization of the
Bethe-Salpeter equation within an all-electron first-principles framework.  We
demonstrate our approach with the emission spectra of the fluorine K edge in
LiF, and provide an in-depth analysis of the pathways that determine the
spectral shape. Excitoninc effects are shown to play a crucial role in both the
valence and core regime.
\end{abstract}

% insert suggested PACS numbers in braces on next line
% insert suggested keywords - APS authors don't need to do this
%\keywords{}

%\maketitle must follow title, authors, abstract, \pacs, and \keywords
\maketitle

%textwidth in inches: \printinunitsof{cm}\prntlen{\textwidth}
%columnwidth in inches: \printinunitsof{in}\prntlen{\columnwidth}
Resonant inelastic x-ray scattering (RIXS) spectroscopy is an important tool to
unravel the nature of elementary excitations. RIXS has been measured to study
excitations in a wide range of crystalline materials
~\cite{Kotani2001,Ament2011,ghiringhelliNiOTestCase2005,
wangChargeTransferAnalysis2p3d2017} and
molecules~\cite{cesarResonantXrayScattering1997,henniesDynamicInterpretationResonant2007,
josefssonInitioCalculationsXray2012}.  Theoretically, the microscopic RIXS
process is commonly described in a two-step model~\cite{Kotani2001,Ament2011}:
In the first step, an incoming x-ray photon with energy $\omega$ is absorbed,
leading to the excitation of a tightly bound core electron to the conduction
band.  Subsequently, a valence electron fills the core hole by emitting an x-ray
photon with smaller energy $\omega'$. The system thus reaches a many-body state
with a hole in a valence state and an excited electron in a conduction state.
This final state is similar to a final state of optical absorption. The two
processes within RIXS occur coherently, \textit{i.e.} the entire process cannot
simply be considered as an absorption followed by an emission
\cite{Ma1992,Ma1994}.  Rather, the final state of the absorption process
determines the possible emission processes. Through this coherence, RIXS
spectroscopy offers an element- and orbital-selective probe of elemental
electronic excitations, because the absorption edge can be selected such to
allow for emission from specific valence states only.

The rich information contained in RIXS spectra has stimulated \textit{ab initio}
descriptions of the microscopic processes. First-principles approaches for RIXS
in solids have been derived with various levels of sophistication,
starting with the independent-particle approximation (IPA) within density
functional theory
(DFT)~\cite{Ma1992,Johnson1994,Jia1996,Strocov2004,Strocov2005}.
Treatment of the electron-hole interaction in the core-hole approximation led to an
improvement ~\cite{Magnuson2010} but still does not capture the full picture.
Electron-hole correlation plays a crucial role in RIXS even in weakly correlated
materials, as both the intermediate and the final excited state of the process
contain an electron-hole pair. This situation demands a more accurate approach
as provided within many-body perturbation theory by the solution of the
Bethe-Salpeter equation
(BSE)\cite{Shirley1998,Shirley2000,Vinson2016,Vinson2017,Geondzhian2018}. This
method is the state-of-the-art for the calculation of optical and x-ray
absorption spectroscopy of condensed matter
\cite{hedi65pr,hybe-loui85prl,Strinati1988,onid+95prl,albr+97prb,bene+98prl,rohl-loui98prl}.
Furthermore, the process requires not only an accurate treatment of both the
intermediate state, which comprises the core hole and an excited electron, as
well as the final state, containing the valence hole and an excited electron,
but also for the the x-ray emission that connects the two many-body states in
the RIXS process.

In this Letter, we propose a novel compact analytical expression for the RIXS
cross section that combines the oscillator strength of the x-ray absorption with
the excitation pathways from all possible intermediate to all possible final
many-body states. These pathways determine how the intermediate many-body states
relax to the final one, and yield an intuitive interpretation of the RIXS
process, while still including  excitonic effects. To derive explicit
expressions for the corresponding oscillator strength and the coherent
emission pathway, we make use of the eigenstates $A_{cv, \lambda}$ of the
BSE Hamiltonian, \textit{i.e.}
$H^{BSE}A_{\lambda}=E_{\lambda}A_{\lambda}$.

\begin{figure}[t!]
\includegraphics[]{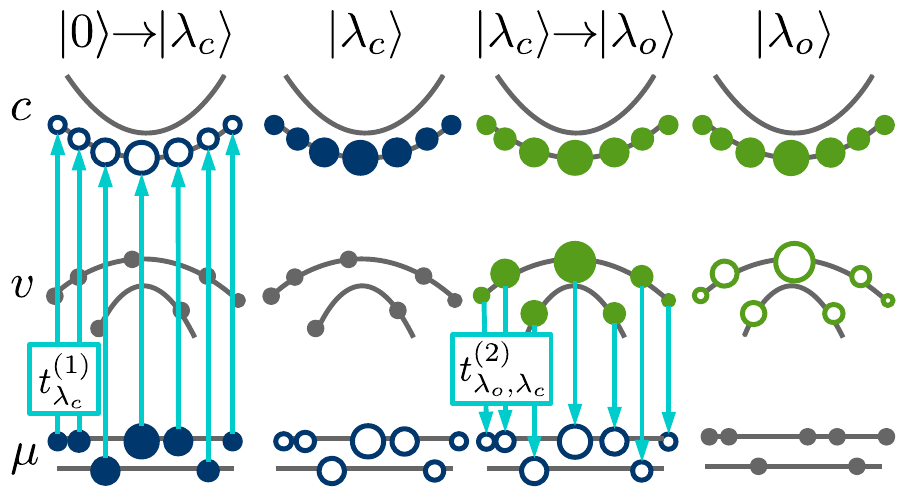}
\caption{\label{fig0:schematics} Scheme of the RIXS process: A core excitation
  yields the intermediate many-body $| \lambda_c \rangle$, where blue circles
  represent the distribution of the core hole (open circles) and the excited
  electron (full circles). Cyan arrows indicate dipole transitions. The
  de-excitation from $| \lambda_c \rangle$ yields the final many-body state $|
  \lambda_o \rangle$. The final state is represented by green circles, where the
  valence hole distribution is shown in open green circles, the distribution of
  the excited electron in full ones.}
\end{figure}
The RIXS process is described by the second-order term in a perturbative
expansion of the electron-photon interaction. The \textit{double
differential cross section} (DDCS) $\frac{d\sigma}{d\Omega d\omega'}$ of
scattering an x-ray photon with energy $\omega$ and polarization
$\mathbf{e}_1$, such that an x-ray photon with energy $\omega'$ and polarization
$\mathbf{e}_2$ is emitted, is given by the Heisenberg-Kramers
formula~\cite{Kramers1925} as
\begin{equation}\label{eq1_1}
  \begin{aligned}
    \frac{\textrm{d}^2 \sigma}{\textrm{d}\Omega\textrm{d}\omega'} \propto 
    \sum_{F}\bigg| \sum_{I} 
    \frac{\langle F |\hat{D}^{\dagger}(\mathbf{e}_2) | I \rangle\langle I
    |\hat{D}(\mathbf{e}_1)| 0 \rangle}{\omega-E_I+\textrm{i}\eta_I}
    \bigg|^2 \times  \\ 
    \times \delta(E_F-E_0+\omega'-\omega),
  \end{aligned}
\end{equation}
where the initial absorption leads to the excitation from the many-body
groundstate $| 0 \rangle$ with energy $E_0$ to an intermediate state $| I
\rangle$ with energy $E_I$. The emission of an x-ray photon leads to the
de-excitation into the final state $| F \rangle$ with energy $E_F$. Here,
$\hat{D}(\mathbf{e})$ describes the dipole transition operator for a given
polarization $\mathbf{e}$. 

Employing the eigenstates $A_{c \mu, \lambda_c}$ with excitation energy
$E^{\lambda_c}$ of the core-level BSE, we define the oscillator strength of core
absorption $t^{(1)}_{\lambda_c}$ as
\begin{equation}\label{eq2_3}
  t^{(1)}_{\lambda_c}=\sum_{c \mu \mathbf{k}} \left[ A_{c \mu \mathbf{k},
    \lambda_c} \right]^* d_{c \mu \mathbf{k}},
\end{equation}
where $c$ are the conduction states of the system, $\mu$ the core states, and
$d_{c \mu \mathbf{k}}(\mathbf{e})=\mathbf{e}\cdot \langle c \mathbf{k}
|\mathbf{p}| \mu \mathbf{k} \rangle$ the momentum matrix elements. With
$t^{(2)}_{\lambda_o,\lambda_c}$ we define the excitation pathway from the
core-level excitation $\lambda_c$ to the valence excitation $\lambda_o$ as
\begin{equation}\label{eq2_5}
  t^{(2)}_{\lambda_o, \lambda_c}=\sum_{cv \mathbf{k}}\sum_{\mu}
  \left[A_{cv\mathbf{k}, \lambda_o}\right]^* d'_{\mu v \mathbf{k}}
  A_{c \mu \mathbf{k},\lambda_c}.
\end{equation}
This definition allows us to finally write the RIXS cross section as
\begin{align}
  \frac{\textrm{d}^2 \sigma}{\textrm{d}\Omega\textrm{d}\omega'}\propto & 
  \mathrm{Im}\sum_{\lambda_o}\frac{\left|\sum_{\lambda_c}
  \frac{t^{(2)}_{\lambda_o,\lambda_c}t^{(1)}_{\lambda_c}}%
  {E^{\lambda_c}-\omega+\mathrm{i}\eta}\right|^2}
  {E^{\lambda_o}-(\omega-\omega')
  -\mathrm{i}\eta}  = \label{eq2_8} \\
  &   = \mathrm{Im}\sum_{\lambda_o}\frac{\left|t^{(3)}_{\lambda_o}(\omega)
  \right|^2}{E^{\lambda_o}-(\omega-\omega')-\mathrm{i}\eta}, \label{eq2_7}
\end{align}
where we have introduced the RIXS oscillator strength
$t^{(3)}_{\lambda_o}(\omega)$.

\begin{figure}[ht]
\includegraphics[]{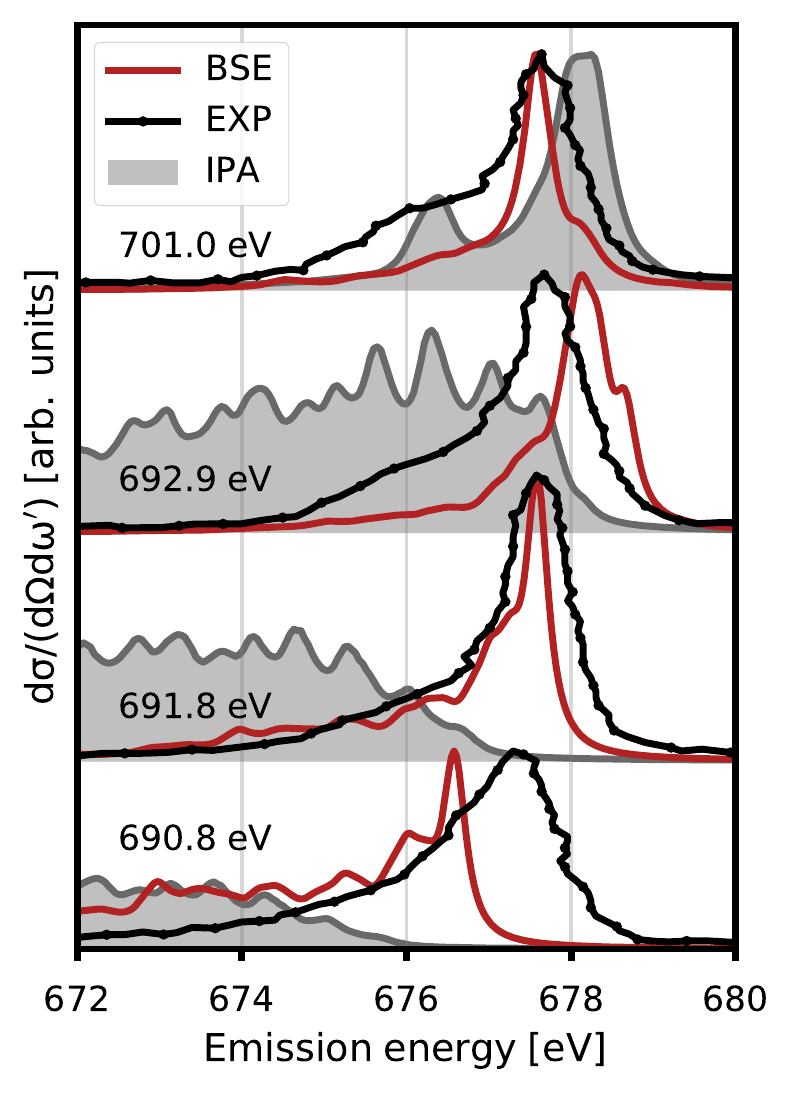}% Here is how to import EPS art
\caption{\label{fig3:LiF-exp}Calculated (red) and experimental
(black)\cite{Kikas2004} RIXS spectrum of the fluorine K edge in LiF. Both
spectra are normalized for each absorption energy. A Lorentzian broadening of
0.15 eV is employed in the calculated spectra.}
\end{figure}
\begin{figure*}[t!]
\includegraphics[]{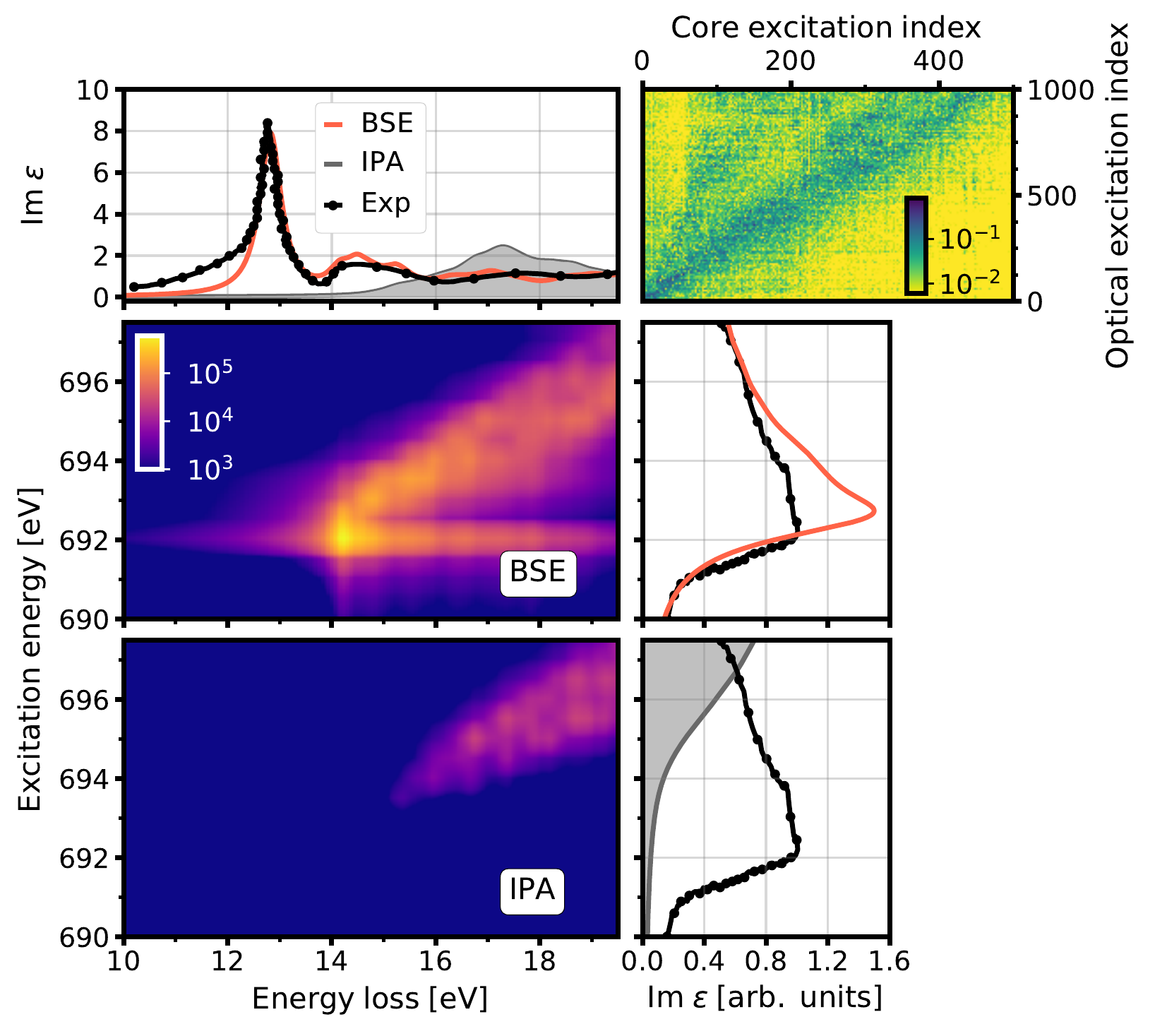}
\caption{\label{fig4:LiF-comb}F K edge RIXS cross section calculated from
  the BSE (center top) and within the IPA (center bottom), together with the
  optical (top left) and the F K edge (right) absorption spectrum. The
  calculated spectra (red BSE, gray IPA) are compared to experimental results
  (black) for the optical~\cite{Rao1975} and core \cite{Joly2017} excitations.
  The excitation pathways $|t^{(2)}|^2$ between the first 500 core and 1000
  optical excitations are shown on the top right.}
\end{figure*}
This compact expression of the DDCS has two advantages: First, it neatly
separates terms that depend on either the excitation energy $\omega$ or the
energy loss $\omega-\omega'$ from those that are independent of energy. This
allows for an efficient numerical evaluation since the most-involved term,
$t^{(2)}_{\lambda_o, \lambda_c}$, is frequency independent. More importantly,
the expression yields an intuitive interpretation of the RIXS spectra in terms
of {\it excitonic pathways} as shown in Fig.~\ref{fig0:schematics}. In essence,
the rate of the initial x-ray absorption event is given by
$t^{(1)}_{\lambda_c}$, together with the energy conservation enforced by the
denominator $E^{\lambda_o}-\omega+i\eta$ in Eq.~\ref{eq2_8}. The absorption
leads to an intermediate state $| \lambda_c \rangle$
\footnote{\label{footnote1}Within the Tamm-Dancoff approximation, the excited
  state is given as $\unexpanded{|\lambda \rangle=\sum_{c\mu \mathbf{k}}A_{c \mu
\mathbf{k},\lambda}\hat{c}^{\dagger}_{c \mathbf{k}}\hat{c}_{\mu \mathbf{k}}| 0
\rangle}$} containing a core hole, schematically shown in
Fig.~\ref{fig0:schematics}.  The final RIXS spectrum is given by the rate of the
first event combined with the pathway $t^{(2)}_{\lambda_o,\lambda_c}$ that
connects the excited state $| \lambda_c \rangle$ with the final state $ |
\lambda_c \rangle$ shown in Fig.~\ref{fig0:schematics}. The pathways
$t^{(2)}_{\lambda_o,\lambda_c}$ depend strongly on the intermediate
state $| \lambda_o \rangle$ due to the coherence of the absorption and emission
processes, and the mixing between $t^{(1)}_{\lambda_c}$ and
$t^{(2)}_{\lambda_o,\lambda_c}$ can result in destructive or constructive
interference, attesting the many-body character of such processes. Another way
to look at the DDCS is provided by Eq.~\ref{eq2_7} which tells that the
overall RIXS signal is given by the combination of the energy loss and the
oscillator strength $t^{(3)}_{\lambda_o}(\omega)$ of the whole process. The
oscillator strength $t^{(3)}_{\lambda_o}(\omega)$ thus solely depends on the
excitation energy, while the dependence on the energy loss is given by the
denominator $E^{\lambda_o}-(\omega-\omega')+\mathrm{i}\eta$. 

Following the equations above, the determintation of the RIXS DDCS
requires the output of two BSE calculations: The first BSE calculation is
performed to obtain the core excitations at a specific edge, which yields
$E^{\lambda_c},A_{c\mu\mathbf{k},\lambda_c}$ as well as the momentum matrix
elements $d_{c \mu \mathbf{k}}$ and $d'_{\mu v \mathbf{k}}$. A second BSE
calculation determines the valence excitations yielding
$E^{\lambda_o},A_{cv\mathbf{k},\lambda_o}$. Subsequently, the absorption
oscillator strength $t^{(1)}_{\lambda_c}$ and the pathway $t^{(2)}_{\lambda_o,
\lambda_c}$ are calculated. Finally, all intermediate quantities are combined to
construct the RIXS oscillator strength $t^{(3)}_{\lambda}(\omega)$ and the DDCS
of Eq.~\ref{eq2_8}. For a consistent treatment of the BSE eigenstates in
the optical and x-ray region, we perform the calculations using the
all-electron many-body implementation in the \texttt{exciting}
code~\cite{Vorwerk2019,Gulans2014}. The all-electron implementation also
directly yields the momentum matrix elements $d_{c \mu \mathbf{k}}$ and
$d'_{\mu v \mathbf{k}}$ between core states and conduction and valence states,
respectively.

To illustrate our approach, we present in the following results for the F K edge
of LiF. Due to its large bandgap, strong effects of electron-hole interaction
are observed, as indicated by the presence of bound excitons in both the valence
and the core regimes. The calculated RIXS spectra as a function of the emission
energy are shown in Fig.~\ref{fig3:LiF-exp} for selected excitation
energies. For an excitation energy of 690.8 eV, below the absorption onset of
approximately 691.8 eV, the calculated spectrum has a peak at 676.8 eV, which
slowly decays at lower emission energies, \textit{i.e.} the maximum of the
scattering occurs at a loss of 14 eV, with considerable contributions at higher
energy loss. With increasing excitation energy, the peak becomes narrower and
moves to slightly higher emission energy. The broad feature at lower emission
energy is strongly suppressed for excitations of approximately 691.8 eV,
while a shoulder in the emission appears for even higher excitations.
The calculated spectra at a given excitation energy, as well as the change as a
function of the excitation energy are in good agreement with their experimental
counterparts \cite{Kikas2004}.

Our approach allows for a deeper analysis of the RIXS spectra, the results of
which are shown in Fig.~\ref{fig4:LiF-comb}. For excitation energies below the
absorption onset of the core edge, i.e., at approximately 691.8 eV, the
cross section is small, since the F $1s$ states are not excited resonantly. This
case is discussed in the Supplementary Information. When the excitation energy
is in resonance with the absorption onset, the spectrum changes abruptly. The
oscillator strength increases tremendously and is shifted to a distinct
loss peak at 14.6 eV. For higher absorption energies, this peak shows a linear
dispersion. It corresponds to the strong emission observed for
absorption energies of 691.8, 692.8 ,and 701 eV in Fig.~\ref{fig3:LiF-exp}.
Furthermore, this feature loses oscillator strength and widens with increasing
excitation energy, thus introducing a shoulder at higher loss with increasing
relative intensity.

As the shape of the RIXS spectrum is determined by the excitation pathways, we
now have a closer look at the $t^{(2)}$-matrix. The top right of
Fig.~\ref{fig4:LiF-comb} shows this matrix for the first 500 core and 1000
valence excitations, which determine the RIXS cross sections for excitation
energies between 680.1 and 696.7 eV and energy losses between 12.8 and 18.3 eV.
It shows a pronounced band-matrix form, \textit{i.e.} the largest contributions
are observed along the diagonal. From Eq.~\ref{eq2_5}, two factors can be
inferred that lead to large pathway matrix elements. First, the
transition from the valence hole distribution of the final state to the core
hole has to be dipole-allowed, and second, the distributions of the excited
electron of the intermediate and final state have to be similar. For core
excitations with increasing energy, the excited electron is distributed farther
from the band gap, and the same holds true for optical excitations with
increasing energy.  This similarity leads to the band-matrix form of $t^{(2)}$.
Moreover, we find that for core excitations at higher energies, pathways to more
and more valence excitations are possible, and therefore the shoulder at higher
loss is getting more pronounced. 

Although the elements  of $t^{(2)}$ yield insight into the origin of the
features in the RIXS spectrum, they do not solely define it. While the pathway
between the lowest excitations in the optical and core spectrum is very strong,
surprisingly, the excitonic peak that dominates the optical absorption spectrum
at 12.7 eV is not observed in the RIXS spectrum. This strongly bound exciton is
formed by a complicate interplay of the bottom of the conduction band, dominated
by the Li $s$ states, and the top of the valence band, formed by the F $p$
states.\cite{gatt-sott13prb} In the corresponding RIXS spectrum, the core
excitation into the Li $s$ states at the bottom of the conduction band is not
possible, as $s \rightarrow s$ transitions are dipole-forbidden. While the
$t^{(2)}$ matrix element between the dark exciton in the F K edge and the bound
exciton in the optical spectrum is considerable, the $t^{(1)}$ entries are zero,
as the initial excitation of the dark exciton is prohibited.

Finally, we demonstrate the importance of electron-hole interaction by comparing
in Fig.~\ref{fig3:LiF-exp} the RIXS spectra obtained by the BSE with those from
the independent-particle approximation (IPA). For low excitation energies, a
broad emission spectrum is predicted within the IPA, missing the pronounced peak
found in both the experimental spectra and our BSE calculations. At an
excitation energy of 701 eV, the agreement between the IPA and BSE
spectra improves, because the effect of electron-hole interaction decreases
with increasing excitation energy. Comparing in Fig.~\ref{fig4:LiF-comb} the
RIXS cross sections obtained from the two calculations, one notices that
the strong peak at the excitation energy of 691.8 eV and the loss of 14.8 eV is
completely missing within the IPA. This comparison demonstrates that the
renormalization of the RIXS spectra at low excitation energies and low energy
loss due to electron-hole interaction is considerable, which contributes
the dominant feature in the RIXS cross section. We note that in the literature,
this peak has been ascribed to an excitonic peak~\cite{Kikas2004}. Our
first-principles approach shows that RIXS spectrum at the core onset is
more complex and requires in-depth analysis to be unraveled.

In conclusion, we have presented in this Letter a novel many-body expression for
the RIXS cross section in condensed-matter systems. Our all-electron
full-potential approach to RIXS treats the electron-hole interaction in
both core and valence excitations consistently, and we show that this careful
treatment of this interaction is paramount to accurately predict
the RIXS cross section. Our results for the wide-gap insulator LiF are in good
agreement with experiment. Importantly, our approach represents a new powerful
analysis tool that allows us to track spectral features back to the coherent
pathways between core and valence excitations, a crucial step to explain and
interpret the richness of RIXS spectra.

This work was supported by STSM grants of the COST Action MP1306 and COST
Action CA17126.

% Create the reference section using BibTeX:

%%%%%%%%%%%%%%%%%%%%%%%%%%%%%%
\end{document}